\documentclass[showpacs,notitlepage,prl,aps,longbibliography,twocolumn,nofootinbib]{revtex4-1}
\usepackage{graphicx}\usepackage{amsmath,amssymb}\usepackage{slashed}

\newcommand{\be}{\begin{equation}}\newcommand{\ee}{\end{equation}}
\newcommand{\bea}{\begin{eqnarray}}\newcommand{\eea}{\end{eqnarray}}

\newcommand{\Ga}{\varkappa}

\renewcommand{\phi}{\varphi}
\def\Perp{{\scriptscriptstyle \| }}
\newcommand{\p}{p_{\Perp}}

\def\bp{\mathbf{p}_\Perp}
\def\bq{\mathbf{q}_\Perp}

\def\({\left(}
\def\){\right)}
\def\[{\left[}
\def\]{\right]}

\def\={\mathop{=}}
\def\seq{\mathop{\simeq}}
\renewcommand\Re{\mathop{\rm Re}}
\newcommand{\xx}{{\rm xx}}
\newcommand{\Ref}[1]{(\ref{#1})}

\renewcommand{\v}{v_{\rm F}}

\def\s#1{\slashed{#1}}
\def\g{\gamma}
\usepackage{color}

\usepackage{ulem}%
\begin{document}
\title{Enhanced Casimir effect for doped graphene
}
\author{M. Bordag${}^1$, I. Fialkovskiy${}^{2,}$\thanks{ifialk@gmail.com}, D. Vassilevich${}^2$}
\affiliation{${}^1$ Leipzig University, Institute for Theoretical Physics,  04109 Leipzig, Germany.\\
${}^2$ CMCC-Universidade Federal do ABC, Santo Andr\'e, S.P., Brazil}
\date{\small \today}

\begin{abstract}
 We analyze the Casimir interaction of doped graphene. To this end we derive a simple expression for the finite
temperature polarization tensor with a chemical potential. It is found that doping leads to a strong enhancement
of the Casimir force reaching almost $60\%$ in quite realistic situations. This result should be important for
planning and interpreting the Casimir measurements, especially taking into account that the Casimir interaction
of undoped graphene is rather weak.
\end{abstract}

\maketitle
\paragraph{Introduction}
Graphene, which is a two-dimensional sheet of carbon atoms possesses many unusual properties and attracts a lot of
attention. Particular excitement among the theoreticians is caused by the fact that the spectrum of quasi-particles
in graphene is described by the quasi-relativistic Dirac model with the effective propagation speed of about 300
times less than the speed of light. 
This continuous model turned out to be very successful in describing a broad range of effects \cite{Katsnelson}, for instance optical properties
of graphene as the absorption of light \cite{abs} and the (giant) Faraday effect \cite{Faraday}, to mention a few.

In the recent years, the Casimir effect for pristine graphene was studied both for   zero \cite{Bordag:2009fz,DW}  and
finite \cite{Dobson,Gomez,Fialkovsky:2011pu} temperatures. For not too large temperatures (as compared with inverse
distance between the interaction sheets) the effect between graphene monolayer and ideal metal is defined by the fine
structure constant $\alpha\simeq 1/137$ and is roughly $2.5\%$
of the one between two ideal metal plates. Such small forces are on the limit of sensitivity of modern experimental
techniques. For high temperatures (or separations) the effect is hugely reinforced \cite{Fialkovsky:2011pu},
but the measurements under such conditions is a separate quite challenging task, which is not completely solved
yet even for metals, see e.g. \cite{Mostepanenko15}. It is not surprising therefore that just a single experiment
has been performed until now \cite{graCas-exp}. This experiment revealed \cite{graCas-compar} a good agreement
with the theory \cite{Bordag:2012hi}. Possibilities, opened by doping, were however not explored there.

Previously the Casimir interaction of doped graphene was studied in \cite{Bo,Sara}. The reflection coefficients used
in that works were expressed though quantities whose explicit dependence on the temperature remained unknown and the results of Refs.\ \cite{Bo,Sara} are mutually contradicting. Therefore, specifically after
experimental confirmation of the Dirac approach to Casimir energy of undoped graphene \cite{graCas-exp,graCas-compar}, it seems to be important to extend the approach to   doped graphene.

In this letter we consider the Casimir effect at finite temperature, chemical potential and mass, and find a substantial enhancement of the   effect in graphene-metal systems which potentially permits to avoid the above mentioned experimental difficulties.  Our findings show that for relatively highly (but still feasibly) doped  graphene monolayers the effect gets stronger by approximately $60\%$. In the (formal) limit of an infinite chemical potential the Casimir interaction becomes $1/2$ of that for the ideal metal. Our calculation is based on a complete representation of the polarization tensor of the fermionic quasi-particles in graphene at finite temperature, chemical potential and mass gap. The obtained result is surprisingly simple and can be easily analytically continued to whole complex frequency plane, including, importantly, real optical frequencies.
It is based on the QED formalism applied to graphene-like systems in \cite{Gusynin,Pyatkovskiy},
and generalizes the results of  \cite{Bordag:2009fz,Fialkovsky:2011pu,Bordag:2015gla}
and those of \cite{Bordag:2015}, which also allows continuation to the whole plane of complex frequencies, to the case of simultaneous presence of  finite chemical potential, finite temperature and non-zero mass-gap.
We also describe a very precise approximation scheme that considerably simplifies the computation of Casimir interaction with graphene.

\paragraph{Model}
The theoretical description of the electronic properties of graphene based on the continuous Dirac model with a   $2+1$-dimensional action.  In the notations of \cite{gra-QFT} it reads
\begin{equation}
	S_{\rm D}=\int d^3x \, \bar\psi (\tilde \gamma^l
		(i\partial_l-eA_l)-m)\psi, \label{Di}
\end{equation}
where $l=0,1,2$ and $x=(x^0,x^1,x^2)$. The gamma matrices $\tilde\gamma^l$ are rescaled,
$\tilde\gamma^0\equiv\gamma^0,\ \tilde\gamma^{1,2}\equiv
v_F\gamma^{1,2},\ \gamma_0^2=-(\gamma^1)^2=-(\gamma^2)^2=1$.
We use natural units $\hbar=c=k_B=1$, and the Fermi velocity is $v_F\simeq
(300)^{-1}$.  Thereby we assume  that the graphene monolayer is placed at the $(x^1, x^2)$ plane. The electromagnetic potential  $A_\mu$ is normalized in such a way that
$e^2\equiv 4\pi\alpha =\frac{4\pi}{137}$.

Reflecting the spin and valley degeneracy in graphene, the gamma matrices, $\g^l$, are $8\times 8$ being a direct
sum of four $2\times 2$ representations (with two copies of each of the two inequivalent ones). The value of the mass gap parameter $m$ and mechanisms of its generation are under discussion \cite{Gusynin,Pyatkovskiy}.

As shown in many previous works, see reference in \cite{gra-QFT}, the electronic properties of graphene in the  formalism of Quantum Field Theory (QFT) can be described by the one loop polarization operator, which is gauge invariant in QED-like model defined by \Ref{Di}. In Minkowski momentum space it is given by
\be
\Pi^{mn}(p) =
	ie^2 \int\frac{d^3 q}{(2\pi)^3}\,\, {\rm tr} \( \hat S(q)\tilde\gamma^m  \hat S(q-p)\tilde\gamma^n\), \label{Pi_expl}
\ee
where $p=(p_0, p_1,p_2)$, $q=(q_0, q_1,q_2)$, and $\hat S$ is the causal (Feynman) propagator of the quasiparticles in graphene, \be \hat S (q_0,\bq) = -\frac{(q_0+\mu)\gamma_0-v_F\s{\bq}-m}{(q_0+\mu+i\epsilon q_0)^2-v_F^2\bq^2-m^2} \label{hat S} \ee ($\epsilon>0$).
Note that due to the quasi-relativistic nature of excitations in graphene $\hat S$ also depends on the Fermi velocity $v_F $.
Further  notations  are $\bq=(q^1,q^2)$, $\slashed{\bq}=\gamma^1q_1+\gamma^2q_2$, and $\mu$ is the chemical potential.

Temperature is introduced using the Matsubara formalism.
In the $\g$-trace in \Ref{Pi_expl}, which can be calculated immediately, see e.g. Eq.\ (A20) in  \cite{Bordag:2015gla},  one has to substitute the integral by a sum,
\be
 i \int dq_0 \rightarrow - 2 \pi T \sum_{k=-\infty}^{\infty}, \quad
        q_0  \rightarrow 2\pi i T(k+1/2)  ,
        \label{Mats_prescr}
\ee
where $k$ is integer. The external frequency of the polarization operator is bosonic,
$p_0  \rightarrow i p_4=2\pi i T n $, $n=0,1,2,\ldots$

\paragraph{Calculation of the polarization operator for finite temperature, mass gap and chemical potential~}
All components of the polarization tensor can be expressed via two scalar quantities (form factors), $\Pi_{\rm tr}\equiv \Pi_{00}-\Pi_{11}-\Pi_{22}$ and $\Pi_{00}$ \cite{Zeitlin,Fialkovsky:2011pu}. As in the QED/QCD cases  these quantities consist of the vacuum part and a part carrying the dependence on $T$ and $\mu$,
\be \Pi_\xx(p;\mu, T) 	= \Pi_\xx^{(vac)}(p) 		+\Delta \Pi_\xx (p;\mu,T), 		\label{decomp1} \ee
where $\xx$ stands either for '${\rm tr}$' or '$00$'. The vacuum part, $\Pi_\xx^{(vac)}(p)$, corresponds to $\mu=T=0$. While such decomposition is a well known feature of polarization tensor in different theories, see e.g. \cite{shuryak}, its realization in particular cases and derivation of simple transparent formulas may be a challenging task. One transforms the sum over the Matsubara frequencies  \Ref{Mats_prescr} into a contour integral consisting of three parts, one of which gives the original integral over the continuous (Euclidean) momenta $q_4$ and the other two can be taken explicitly by the Cauchy theorem. The remaining integral over the in-plane momenta, $\bq=(q_1,q_2)$, can be further simplified by performing the angular integration. Recently such procedure was applied to graphene at $\mu=0$, $T\ne0$ in \cite{Bordag:2015} and $T=0$, $\mu\ne0$ in \cite{Bordag:2015gla}.
Omitting the technicalities we arrive at
\begin{eqnarray}
&&
\Delta\Pi_\xx =\label{Delta Pi}\\
&&
\frac{ 8\alpha }{v_F^2}
	\int_m^\infty d\varkappa
	\(1+ \Re\frac{M_\xx }{\sqrt{Q^2-4\p^2 (\Ga^2-m^2)}} \) \Xi (\varkappa) .
	\nonumber
\end{eqnarray}
Here the distribution function, $\Xi\equiv {(e^{(\Ga+\mu)/T}+1)^{-1}} + {(e^{(\Ga-\mu)/T}+1)^{-1}}$, carries the dependence on $T$ and $\mu$. Further notations in \Ref{Delta Pi} are
\begin{eqnarray} &&M_{00}	= -\tilde p^2+4i p_4 \Ga+4\Ga^2 , \nonumber\\ && M_{\rm tr}=  -\tilde p^2+ 4 \Ga (1-\v^2) (i p_4 + \Ga) + 4 \v^2 m^2,\nonumber\\ &&Q=\tilde p^2 -2i p_4 \Ga,\quad 	\tilde p^2 \equiv p_4^2+v_F^2\p^2, \quad 		\p=|\bp|. \nonumber \end{eqnarray}
Note that $\Delta	\Pi_\xx$, \Ref{Delta Pi}, does not have UV singularities.

For the vacuum part, $\Pi^{\rm (vac)}$, one can directly use the well-known expressions \cite{Bordag:2009fz} valid for graphene,
\begin{equation}
	\Pi^{\rm (vac)}_{00}=\frac{\alpha\Phi \p^2}{\tilde p^2} \,,\qquad
	\Pi^{\rm (vac)}_{\rm tr}=\frac{\alpha \Phi (p^2+\tilde p^2)}{\tilde p^2}
\,,\label{Pivac}
\end{equation}
where 
$
\Phi=4 \left[ m + \frac{\tilde p^2 -4m^2}{2\tilde p} 
\arctan \left(\frac{\tilde{p}}{2m}\right)\right]
$.
From now on we set $m=0$ (gapless graphene) unless otherwise stated.

One of the advantages of the decomposition \Ref{decomp1} is the absence of the summation over Matsubara frequencies, which permits relatively easy derivation of the limiting cases. In particular, in the limit of zero temperature (but not zero chemical potential) we obtain
\begin{eqnarray}
	&&\Delta	\Pi_{tr} ={8\alpha} \times \label{t0-1} \\
	&&   \(
	\frac\mu{v_F^2}
	- {\rm Im}\[
	    \frac{\tilde p^2+p^2}{4 \tilde p } \log\(x+\sqrt{x^2+1}\)
	    \right.
	\right.
\nonumber \\
&&
  \left.\left.\qquad+\frac{(1-v_F^2)\p^2}{4\tilde{p}}  x\sqrt{x^2+1}\]\)\nonumber\\
	&&\Delta	\Pi_{00} =
	  {8\alpha }   \times\label{t0-2} \\
&& \(
		\frac\mu{v_F^2}
		-  \frac{\p^2}{4 \tilde p}{\rm Im} \[
			x\sqrt{x^2+1}
			+ \log\(x+\sqrt{x^2+1}\)
		\]
	\),
\nonumber
\end{eqnarray}
with $ x=\frac{2 i \mu- p_4}{v_F \p}$. The formulae above are the analogue of the (B2) \cite{Bordag:2015gla} taken at Matsubara frequencies, and one can check that in the appropriate limits they reproduce the results of other authors \cite{Bordag:2009fz,Fialkovsky:2011pu,Gusynin,Pyatkovskiy,Bordag:2015, Bordag:2015gla}.
Similar to the results of \cite{Bordag:2015}, the representation \Ref{decomp1} with \Ref{Delta Pi} directly permits for continuation to the real frequencies, and thus can be applied for investigation of the optical properties, surface plasmons and other effects in graphene for finite temperature and chemical potential.

\paragraph{Enhancement of the Casimir effect} The Casimir energy density (per unit area) for two parallel interfaces separated by the distance $a$ is given by the Lifshitz formula
\cite{Lifshtz} in terms of the reflection coefficients,  $r^{(1)}_{\rm TE, \rm TM}$, $r^{(2)}_{\rm TE, \rm TM}$, of the TE and TM electromagnetic modes on the two interfaces,
\begin{equation}
    {\mathcal E}
    =k_BT\sum_{n=-\infty}^\infty\int\frac{d^2\bp}{8\pi^2}
	\sum_{X=TE,TM} \ln (1-e^{-2p_\| a}r_{\rm X}^{(1)}r_{\rm X}^{(2)}),
        \label{EL}
\end{equation}
where $p=\sqrt{p_4^2+\p^2}$, and $p_4=2\pi n T$ are the
Matsubara frequencies 
\footnote{To restore physical units in Eq.\ (\ref{EL}) it is enough to apply the conversion rule
$({\rm eV})^{-1} \sim 1.97\cdot 10^{-7}{\rm m}$.}. 
The reflection coefficients are
taken at Euclidean momenta $r=r(p_4,\bp)$.
They were derived in \cite{Fialkovsky:2011pu} in terms of the polarization operator components $ \Pi_{00, \rm tr}$,
\be
r_{\rm TM}=\frac{p   \Pi_{00}}{p   \Pi_{00}  + 2  \p^2},  \quad
r_{\rm TE}= - \frac{ p ^2  \Pi_{00}- \p^2  \Pi_{\rm tr}}
            {p^2 \Pi_{00} -  \p^2 ( \Pi_{\rm tr} +2  p )},
\label{rTETM-grPi}
\ee
 (and rederived in numerous papers afterwards).
For the perfect conductor,
$  r_{\rm  TM}^{(2)}=1,\   r_{\rm  TE}^{(2)}=-1$.
Combining \Ref{EL} with (\ref{rTETM-grPi}) and using (\ref{decomp1},\ref{Delta Pi}) for the polarization operator at finite temperature and chemical potential, we are able to calculate the Casimir energy density, $\cal E$, the Casimir pressure ${\mathcal F}=-\partial {\cal E}/\partial a$, and its' gradient,  ${\mathcal G}=\partial {\cal F}/\partial a\equiv -\partial^2 {\cal E}/\partial a^2$,  between a doped graphene layer and an ideal metal plate.

It is instructive to consider the case of very large $\mu$ first. In the formal limit $\mu\to\infty$, both $\Pi_{00}$ and $\Pi_{\rm tr}$ have identical asymptotics,
\be
	\Pi_{\xx}\seq_{\mu\to\infty}  \frac{8\alpha }{v_F^2}\mu+\ldots,
	\label{limit}
\ee
which can be readily deduced from (\ref{t0-1},\ref{t0-2}). Thus the electronic properties of graphene would be expected to become closer to those of ideal metal. However, due to the specific structure of the reflection coefficients \Ref{rTETM-grPi}, it is only the contribution of TM mode   to the Casimir interaction which grows in this limit.  Thus, at $\mu\to \infty$, the Casimir interaction reaches the value of $1/2$ of the ideal metal - ideal metal one,
\be
   {\mathcal{E}} \=_{\mu\to\infty}
       \frac12 {\mathcal E}_{\rm id}
	= -\frac{k_BT \zeta(3)}{16 \pi a^2}.
	\label{1halve}
\ee
The same result was obtained for the high temperature limit \cite{Fialkovsky:2011pu}. Interestingly, the Casimir interaction of graphene can never reach $100\%$ of that for the ideal metal. Eq.\ (\ref{1halve}) gives a very rough idea on how far the enhancement of the Casimir effect with $\mu$ might go. Practically, it hardly makes sense to consider $|\mu|$ exceeding a couple of eV in the framework of the Dirac model.

\begin{figure}
\centering \includegraphics[width=8cm]{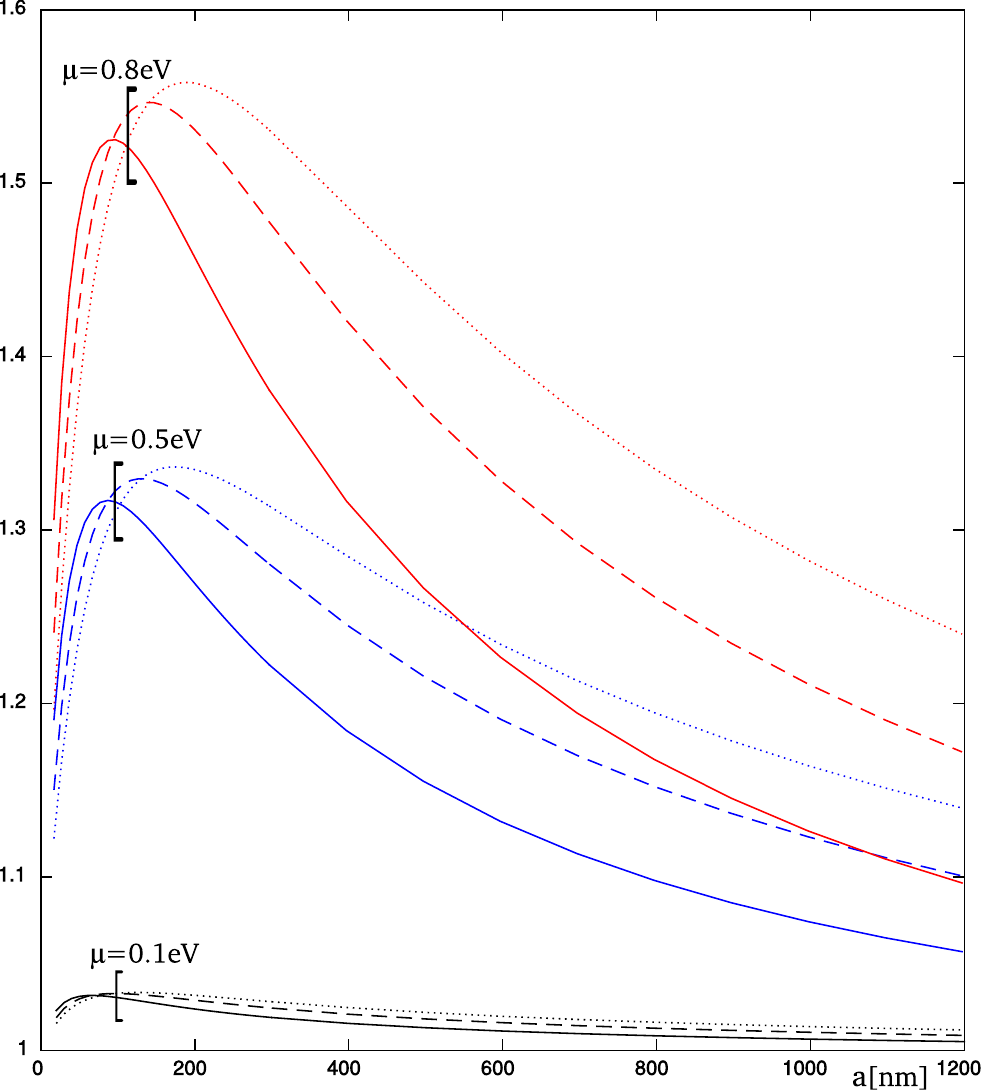}
\caption{
The ratios of Casimir energy density (full lines), pressure (dashed lines) and its' gradient (dotted lines) at $\mu=0.1,0.5,0.8$eV (black, blue and red lines, correspondingly, colour online) and at $\mu=0$, between a perfect metal plate and doped graphene, as a function of distance $a$, [nm].
\label{fig}}
\end{figure}

Our numerical analysis show that at distances about $100$--$300$ nanometers the Casimir effect between a perfect metal plate and doped graphene is highly enhanced even for relatively moderate values of the chemical potential.
On Fig.~\ref{fig} we compare the behavior of the ratios of the energy, pressure and the pressure gradient at given values of the chemical potential to the corresponding values for pristine graphene, as functions of distance.
For $\mu=0.1$eV, the interaction force for the case of doped graphene is only $3.5$\% higher then in the case of pristine graphene. However, already for $\mu=0.5$eV their ratio has a maximum of $32$\% at approximately $70$nm, and for $\mu=0.8$eV the force between a doped graphene layer and an ideal metal is almost $60$\% higher then that for a pristine one. All our simulations are performed at $T=300$K and $m=0$.

One further notices, that the effect is more pronounced the more derivatives we calculate of the energy. Thus, the
ratio of the energy at $\mu=0.8$eV to its pristine value is $1.52$ at maximum, for the force ${\cal F}$ it is $1.54$ and for the gradient $ {\cal G}$ it reaches $1.56$. Moreover, at larger distances ($400$--$1000$ nm), the enhancement effect for the gradient diminishes much slower then that for the energy, which suggests preference for gradient force experiments.

Reaching the values of  the chemical potential of $0.8$eV and higher might be a challenging task and would require preparation of special samples. Without special treatment, the chemical potential in epitaxial graphene layers stays low up to the level of $0.3$--$0.4$~eV or smaller \cite{coletti10,riedl10}. However, the Fermi energy shifts of order $0.8$ eV are achievable in epitaxial single layer graphene due to molecular doping \cite{Zhou08}.
Under certain circumstances, doping may lead to considerable inhomogeneities in the charge distribution, see e.g. \cite{inhomo},
that may give rise to additional forces in the Casimir experiments. Due to the strong charge density dependence on the nature of acceptor/donor mechanism, these forces should be treated individually for each particular experiment.

The distance dependence of the Casimir energy for doped graphene will be altered as compared to the pristine one as it is usual when an additional dimensionfull parameter is introduced. However, its detailed study  is beyond the scope of the present letter.

\begin{figure}
\centering \includegraphics[width=8cm]{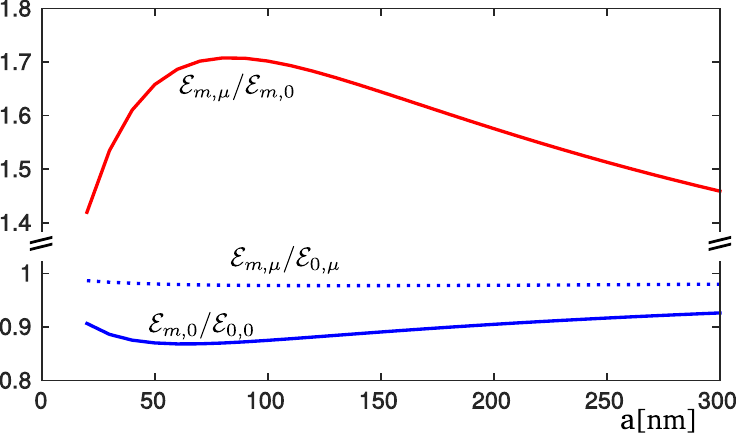}
\caption{
The ratios of Casimir energies between a perfect metal plate and graphene, as a function of distance $a$, [nm]. $m=0.1$eV, $\mu=0.8$eV.
\label{fig-mass}}
\end{figure}

\paragraph{Influence of the mass gap.}
From the physical point of view, the larger is the mass parameter in \Ref{Di} the less should be the conductivity of quasi-particles and, consequently, the smaller the Casimir effect. This was shown explicitly in \cite{Bordag:2009fz} for $T=\mu=0$.
One can also show that the influence of mass is negligible as far as $m\ll\mu$. In particular, in the formal limit \Ref{limit}  any dependence on the mass disappears.
Our numerical simulation shows, see Fig \ref{fig-mass}, that for $\mu=0.8$eV and $m=0.1$eV  doping gives  up to $70$\% enhancement of the Casimir energy density (red line, to be compared with the full red line of Fig. \ref{fig}).  In the same picture it is shown that the influence of the mass gap on the value of the energy for doped graphene (blue dotted line) is almost negligible, while the energy for pristine graphene gets lower by about $15$\% (blue full line). Therefore, doping becomes even more important for gaped graphene.

\begin{figure}
\centering \includegraphics[width=8cm]{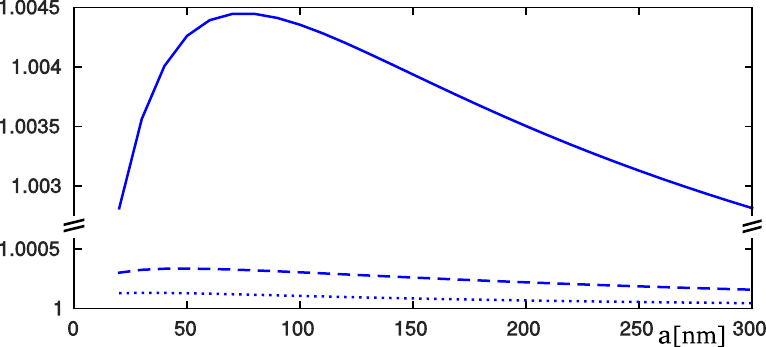}
\caption{
The ratios of Casimir energy to its approximation according to \cite{Fialkovsky:2011pu},
{${\cal E}/{\cal E}_{app}$},
for $\mu=0$, $0.5$, $0.8$eV (full, dashed and dotted lines respectively)
between a perfect metal plate and doped graphene, as a function of distance $a$, [nm].
\label{fig2}}
\end{figure}

\paragraph{Approximating the Casimir energy.}
For finite temperature, the numerical calculation of the Casimir energy in a realistic set-up requires summing   a
large number of contribution to the sum over the Matsubara frequencies. Following the ideas of \cite{Fialkovsky:2011pu} (which
were later confirmed in \cite{KM15}), one might facilitate significantly this calculus by considering the $T=0$
approximation (\ref{t0-1},\ref{t0-2}) for the polarization operator in all terms of the summation \Ref{EL} except in
the zeroth one. The comparison of exact results for the Casimir energy ${\cal E}$ with such approximation, ${\cal E}_{app}$, is given on Fig.~\ref{fig2}. As one can see,
the error is less then $0,5$\% for the pristine graphene and one order of magnitude smaller for doped one. This confirms
once again at an even better level the asymptotic considerations delivered in  \cite{Fialkovsky:2011pu}.

\paragraph{Summary.}
In this letter we calculated the polarization operator for the quasi-particles in graphene at non-zero temperature,
chemical potential and mass gap applicable at all complex frequencies without a need for any special procedure of
analytical continuation. This result can be used in a variety of physical problems, including investigation of TE
surface plasmons in graphene \cite{Bordag:2015gla}, quantum reflection \cite{Dufour13}, Casimir interaction, etc.

Basing on these results and the Lifshitz formula, we numerically simulate the Casimir interaction between a doped
graphene monolayer and an ideal metal. For high but feasible doping we predict the enhancement of the Casimir effect
as compared to the case of the pristine graphene of up to $54$\% for the Casimir force, and of up to $56$\% --- for the force gradient.  The high doping of graphene is shown to bring significant enhancement in the values of the force gradient at a wide range of separations which should facilitate   future experimental measurements.  At such levels of doping the influence of the mass-gap is not important.
We saw also that even moderate values of the chemical potential have a non-negligible effect on the Casimir force and thus
should be taken into account in realistic description of experiments together with real material properties, finite temperature and mass-gap parameter, if present.

All calculations of the present paper were performed in fully retarded approach valid at all distance. It may be interesting to study whether the non-retarded approach may deliver a good approximation at some distances.

Finally we note, that the considerations given in this letter are concerned with the proper graphene properties and the enhancement of the Casimir interaction is invoked by the change in its conductivity. Thus, we can conclude that even in the experiments involving a real metal and/or graphene on a substrate the enhancement effect must be present, though its particular value may differ from the one presented here. The good concordance between the force gradient measurements \cite{graCas-exp} and
theoretical considerations presented in \cite{graCas-compar} shows that the graphene samples used in \cite{graCas-exp} were rather pristine.

\paragraph{Acknowledgments}
This work was supported in part by CNPq and FAPESP.

\end{document}